\documentclass[aps,pra,twocolumn,amsmath,showpacs,amssymb,groupaddresses,superscriptaddress]{revtex4}
\usepackage{times}
\usepackage{latexsym}
\usepackage{graphicx}
\usepackage{amsmath,amssymb}
\usepackage{verbatim,bbm}
\newtheorem{theorem}{Theorem}

\begin{document}

\title{Robust quantum data locking from phase modulation}

\author{Cosmo Lupo}
\affiliation{Research Laboratory of Electronics, Massachusetts Institute of Technology, Cambridge, MA 02139, USA}

\author{Mark M. Wilde}
\affiliation{Hearne Institute for Theoretical Physics, Department of Physics and Astronomy, 
Center for Computation and Technology, Louisiana State University,
Baton Rouge, LA 70803, USA}

\author{Seth Lloyd}
\affiliation{Research Laboratory of Electronics, Massachusetts Institute of Technology, Cambridge, MA 02139, USA}
\affiliation{Department of Mechanical Engineering, Massachusetts Institute of Technology, Cambridge, MA 02139, USA}

\begin{abstract}
Quantum data locking is a uniquely quantum phenomenon that allows a relatively short key  
of constant size to (un)lock an arbitrarily long message encoded in a quantum state, in such
a way that an eavesdropper who measures the state but does not know the key 
has essentially no information about the message.
The application of quantum data locking in cryptography would allow one to overcome 
the limitations of the one-time pad encryption, which requires the key to have the
same length as the message.
However, it is known that the strength of quantum data locking is also its
Achilles heel, as the leakage of a few bits of the key or the message may in
principle allow the eavesdropper to unlock a disproportionate amount of information.
In this paper we show that there exist quantum data locking schemes that can be made robust 
against information leakage by increasing the length of the key by a proportionate amount.
This implies that a constant size key can still lock an arbitrarily long message 
as long as a fraction of it remains secret to the eavesdropper.
Moreover, we greatly simplify the structure of the protocol by proving that 
phase modulation suffices to generate strong locking schemes, paving the way
to optical experimental realizations.
Also, we show that successful data locking protocols can be constructed
using random codewords, which very well could be helpful
in discovering random codes for data locking over noisy quantum channels.
\end{abstract}

\pacs{03.65.-w, 03.67.-a, 03.67.Dd}

\maketitle

\section{Introduction}

Quantum data locking (QDL) is a uniquely quantum protocol that provides one of the strongest 
violations of classical information theory in the quantum setting~\cite{QDL}.
In QDL the knowledge of a relatively short key of constant size allows one to (un)lock an
arbitrarily long message encoded in a quantum system. 
Otherwise, without knowledge of the key, only a negligible amount of information 
about the message can be accessed~\cite{CMP,Buhrman,Leung,Dupuis,Fawzi}.
Such an exponential disproportion between the length of the key and that of the message
is impossible in classical information theory, according to which
the bits of secret key should be at least as many as the bits of encrypted information~\cite{Shannon}.
Although cryptographic applications may seem natural, it was  
recognized early that the security provided by QDL is in general not robust 
under the leakage of a small fraction of the key or the message.
Indeed, as a relatively short key is sufficient to lock a much longer 
message, it may very well happen that the leakage to the eavesdropper 
of a few bits is sufficient to unlock a 
disproportionate amount of information.
Here we analyze this issue and show that there exist QDL schemes that
can be made resilient against information leakage at the cost of 
increasing the length of the secret key by a proportional amount.
We show that to protect QDL from the leakage of $n$ bits of the key 
or the message, it is sufficient to add an overhead of about $n$ bits 
to the secret key.

The security of QDL is based on the {\it accessible information}
criterion. It is well known that such a criterion does not guarantee composable
security~\cite{Renner}.
That is, security is in general not granted if QDL is used as a subroutine of another protocol.
To avoid this problem, one should make an assumption on Eve's technological capability, and
require that the message exchanged by QDL is not used for the next protocol 
until Eve's quantum memory expires.
For instance, composable security is granted if the eavesdropper has no
quantum memory and is hence forced to measure her share of the quantum system
as soon as she obtains it.
Such additional assumptions make the accessible information criterion weaker than the commonly
accepted security criterion for quantum private communication~\cite{Renner}, which is instead
related to the {\it Holevo information}.
Interestingly enough, a large gap exists between these two security criteria that may allow for 
high rate QDL through quantum channels with poor privacy~\cite{QEM,PRX}. 
As a matter of fact, explicit examples of channels with low or even zero privacy that allow
for QDL at high rates have been recently provided~\cite{AW,seplock}.
It is hence of fundamental interest to seek a deeper understanding of the phenomenon of QDL,
especially in the presence of noise.


A number of QDL methods exist.
In some of the best known, the codewords are created by applying
uniformly distributed random unitary operations to the vectors of
a given orthonormal basis~\cite{CMP,Buhrman,Dupuis,Fawzi}.
The role of the random unitaries is to scramble the codewords in such a way that
an eavesdropper has essentially no information about the message.
The crucial feature of strong QDL schemes is that the number of scrambling unitaries
is much smaller (in fact exponentially smaller) than the number of different messages~\cite{NOTA-design}. 
Although a scheme that can be implemented efficiently 
on a quantum computer exists~\cite{Fawzi}, the realization 
of QDL with currently available technologies still presents 
severe experimental difficulties.
Moreover, all known QDL protocols are defined for a noiseless quantum system. Hence,
a problem of fundamental importance is to design protocols capable of performing QDL
through noisy quantum channels. (Explicit protocols for QDL through noisy channels have
been recently introduced in~\cite{seplock}.)


Here we present two QDL protocols based on random phase shifts.
The first one is based on random vectors (instead of random bases) of the form~\cite{NOTA-coding}
\begin{equation}\label{phase-v}
|\psi \rangle = \frac{1}{\sqrt{d}} \sum_{\omega=1}^d e^{i\theta(\omega)} |\omega\rangle \, ,
\end{equation}
where $\{|\omega\rangle\}_{\omega=1,\dots,d}$ is a given orthonormal basis in a $d$-dimensional
Hilbert space, and 
$e^{i\theta(\omega)}$ are i.i.d.\ random phases with zero mean, 
$\mathbb{E}\left[ e^{i\theta(\omega)} \right] = 0$. 
Even the choice $e^{i\theta(\omega)} = (-1)^{b(\omega)}$ is sufficient, where 
$b(\omega)$ are random binary variables. 
We show that codewords sampled from this ``phase ensemble'' of vectors suffice to build strong QDL schemes.
It is worth noticing that, compared to previously known QDL 
protocols that require the preparation of uniform (Haar distributed)
basis vectors, our scheme greatly simplifies the structure of codewords 
for QDL and represents a major simplification for optical experimental 
implementations.

Moreover, the expansion of the set of QDL codewords given in this paper
paves the way to the application of random coding techniques to lock classical
information into noisy quantum channels~\cite{QEM,PRX}, which requires the
codewords used to hide information to coincide with the codewords used to protect 
information from noise.
Codewords randomly selected from an ensemble that attains
the coherent information bound suffice to protect information from
noise~\cite{Lloyd,Shor,Devetak,PSW,Horo,Klesse}.
This paper shows that such codewords also suffice
to keep information secure from an eavesdropper.


The paper proceeds as follows.
In Sec.~\ref{sec:phase} we describe a new QDL protocol where the codewords are obtained
by applying a random phase modulation. Section~\ref{sec:sketch} provides a sketch of the proof
of the QDL property of our protocol, while details are provided in the Appendix~\ref{app:proof}.
Then, Sec.~\ref{sec:robust} proves the robustness of QDL to loss of information to the 
eavesdropper. Section~\ref{sec:appl} discusses possible applications and experimental
realizations of our protocol in quantum optics. Finally, conclusions are presented in
Sec.~\ref{sec:concl}.


\section{Quantum data locking from phase modulation}\label{sec:phase}

In a typical QDL protocol, the legitimate parties, Alice and Bob, publicly
agree on a set of $n = MK$ codewords in a $d$-dimensional quantum system.  
From this set, they then use a short shared private key of $\log K$ bits to select a set
of $M$ codewords that they will use for sending information.
If an eavesdropper Eve does not know the private key, then the number of bits 
--- as quantified by the {\it accessible information} $I_{\text{acc}}$, which is
defined as the maximum mutual information between the message and Eve's measurement result --- 
that she can obtain about the message by intercepting and
measuring the state sent by Alice is essentially equal to zero
for certain choices of codewords.

We consider here two QDL protocols where Alice and Bob are able to communicate 
via a $d$-dimensional noiseless quantum channel.
In the first QDL protocol, to encrypt a message $m \in \{ 1, \dots, M \}$, Alice prepares one of the vectors 
\begin{equation}\label{codews}
|\psi_{mk}\rangle = \frac{1}{\sqrt{d}} \sum_{\omega=1}^d e^{i\theta_{mk}(\omega)} |\omega\rangle \, ,
\end{equation}
where the value of $k \in \{ 1, \dots, K \}$ is determined by the secret key, and
the vectors are sampled i.i.d.\ from the phase ensemble defined above.

We require that Bob, knowing the key, is able to decode with a probability
of success close to $1$.
That is, for any $k$ there exists a positive operator-valued measurement
(POVM) with elements $\{ \Lambda^{(k)}_m \}$ such that
\begin{equation}
\bar{p}_\mathrm{succ} = \frac{1}{M} \sum_{m=1}^M \mathrm{Tr} \left( \Lambda^{(k)}_m |\psi_{mk}\rangle\langle\psi_{mk}| \right) \geq 1 - \epsilon \, .
\end{equation}

On the other hand, we require that if Eve intercepts and measures the state $|\psi_{mk}\rangle$, 
then she will only be able to access a negligible amount of information about the message $m$,
as quantified by the accessible information $I_{\text{acc}}$. 
We require that~\cite{NOTA1}
\begin{equation}
I_{\text{acc}} \lesssim \delta \log{M} \, .
\end{equation}
Furthermore, for a key-efficient QDL scheme we demand that $K \ll M$. 
In particular, we require that $\log{K} = O\left( \log{\log{M}} \right)$
and that $\delta \to 0$ as $K$ increases.


Here we show that a set of $MK$ codewords randomly selected from the phase ensemble will define a
QDL protocol with probability arbitrarily close to $1$ for $d$ large enough.
To show that, we make repeated applications of the following bound on the largest and smallest
eigenvalues of a random matrix:

\begin{theorem}\label{ThRM}~\cite{BaiYin}
Consider a $d \times n$ matrix $W$, 
whose entries are independent and identically distributed random
variables with zero mean, variance $\sigma^2$, and finite fourth moment.
Define $X = (1/n) W W^\dagger$.  Then almost surely 
as $d \rightarrow \infty$, the largest eigenvalue of $X$
$\rightarrow (1+\sqrt y)^2 \sigma^2$, where $y=d/n$.
In addition, when $d \leq n$, the smallest eigenvalue of $X$ $\rightarrow
(1-\sqrt y)^2 \sigma^2$.
\end{theorem}
To apply this theorem to our case, let 
\begin{equation}
W = \sum_{j=1}^n | \psi_j \rangle \langle e_j | \, ,
\end{equation}
where $| \psi_j \rangle$ are $n$ random vectors from the phase ensemble, and
the set $\{|e_j\rangle \}$ is an orthonormal basis for an auxiliary $n$-dimensional Hilbert space.  
Notice that the elements of $W$ are just the components 
of the randomly selected codewords $|\psi_j\rangle$ in the basis $\{ |\omega\rangle\}_{\omega=1,\dots,d}$.  
We have 
\begin{equation}
X = \frac{1}{n} W W^\dagger = \frac{1}{n} \sum_{j=1}^n |\psi_j\rangle\langle \psi_j| \, ,
\end{equation}
and for the phase ensemble $\sigma^2=1/d$.


For finite $d$, we use another result from random matrix theory:
\begin{theorem}\label{ThRM2}~\cite{FS}
The probability that the largest eigenvalue of $X$ is larger than
$\left[ \left( 1 + \sqrt{y} \right)^2 + \delta \right] \sigma^2$ 
is no greater than $C \exp{(-d \delta^{3/2}/C)}$, where $C$ is a constant.
Similarly, if $d < n$, the probability that the smallest eigenvalue is less than 
$\left[ \left( 1 - \sqrt{y} \right)^2 - \delta \right] \sigma^2$ is less than
$C/\left(1-\sqrt{y}\right) \exp{(-d \delta^{3/2}/C)}$.
\end{theorem}
This theorem states that the probability that the bounds of Theorem~\ref{ThRM} are violated by
more than $\delta$ vanishes exponentially with $d$ and $\delta$.


First of all, using these results, we can easily show that for $M \ll d$ Bob's average probability of successful 
decoding is $\bar{p}_\mathrm{succ} \gtrsim 1-2\sqrt{M/d}$.
To see that, assume that Bob applies the ``pretty good measurement'' POVM with elements
\begin{equation}
\Lambda^{(k)}_m = \Sigma_k^{-1/2} |\psi_{mk} \rangle \langle \psi_{mk}| \Sigma_k^{-1/2} \, ,
\end{equation}
where
\begin{equation}
\Sigma_k = \sum_{m=1}^M |\psi_{mk} \rangle \langle \psi_{mk}| \, .
\end{equation}
Then we have, assuming $\delta \ll 1$,
\begin{eqnarray}
\bar{p}_\mathrm{succ} 
          & = & \frac{1}{M} \sum_{m=1}^M \left| \langle \psi_{mk}| \Sigma_k^{-1/2} |\psi_{mk} \rangle \right|^2 \\
          & \geq & \frac{d}{M} \left[ \left( 1 + \sqrt{\frac{d}{M}} \right)^{2} + \delta \right]^{-1} \label{ineq} \\
          & \simeq & 1 - 2 \sqrt{\frac{M}{d}} \, ,
\end{eqnarray}
where in~(\ref{ineq}) we have applied Theorems~\ref{ThRM} and~\ref{ThRM2} to bound
\begin{equation}
\Sigma_k^{-1/2} \geq  \sqrt{\frac{d}{M}} \left[ \left( 1 +  \sqrt{\frac{d}{M}} \right)^2 + \delta \right]^{-1/2} \, .
\end{equation}


On the other hand, the bound on Eve's accessible information is given by the following
\begin{theorem}\label{main-r}
Select $MK$ i.i.d.\ random codewords $|\psi_{mk}\rangle$ ($m=1,\dots,M$ and $k=1,\dots, K$)
from the phase ensemble, with $MK \gg d$ and $M \ll d$. 
Then, for any $\delta >0$ and
\begin{equation}\label{K-condition}
K > \frac{4}{\delta^2} \left( \ln{M} + \frac{2d}{\delta M} \ln{ \frac{5}{\delta} }\right) \, ,
\end{equation}
Eve's accessible information satisfies
\begin{equation}
I_{\text{acc}} = O\left( \delta \log{d} \right)
\end{equation}
up to a probability 
\begin{equation}\label{p-fail}
p_\mathrm{fail} \leq \exp{\left[ - M \left( \frac{\delta^3 K}{4} - \delta \ln{M} - \frac{2d}{M} \ln{\frac{5}{\delta}} \right) \right]}
\end{equation}
that vanishes exponentially in $M$.
\end{theorem}
The sketch of the proof is provided in Sec.~\ref{sec:sketch}, while details
are in Appendix~\ref{app:proof}.

This theorem states that a set of $MK$ random codewords from the phase ensemble defines
a strong QDL protocol.
For instance, if we put $\delta \simeq 1/\log{M}$, then $I_{\text{acc}}$ is smaller than a constant with
$\log{K} = O\left( \log{\log{M}} \right)$. 
Otherwise, for $\delta \ll 1/\log{d}$, a secret key of size $\log{K} = O\left( \log{1/\delta} \right)$
is sufficient to guarantee $I_{\text{acc}} = O\left( \delta \log{d} \right)$.


\subsection{Quantum data locking from random unitaries}\label{ssec:unitary}

The second QDL protocol is defined by a set of random unitaries of a particular form.
We define the ``phase ensemble'' of unitaries of the form
\begin{equation}
U = \sum_{\omega=1}^d e^{i \theta(\omega)} |\omega\rangle\langle\omega| \, ,
\end{equation}
where $\theta(\omega)$ are i.i.d.\ random phases with $\mathbb{E}[e^{i \theta(\omega)}] = 0$.
Given the set of Fourier-transformed basis vectors
\begin{equation}\label{phase-U-q}
|m\rangle = \frac{1}{\sqrt{d}} \sum_{\omega=1}^d e^{i 2\pi m \omega/d} |\omega\rangle \, ,
\end{equation}
for $m=1,\dots,d$, we define a set of $dK$ codewords as
\begin{equation}\label{phase-U}
|\psi_{mk}\rangle = U_k |m\rangle = \frac{1}{\sqrt{d}} \sum_{\omega=1}^d e^{i 2\pi m \omega/d + i \theta_k(\omega)} |\omega\rangle \, .
\end{equation}

Notice that for any $k$, codewords with different $m$ are mutually orthogonal.
This implies that Bob can decode with $\bar{p}_\mathrm{succ} = 1$.
Furthermore, each codeword in~(\ref{phase-U-q}) has the same distribution
of the codewords selected from the phase ensemble of vectors. The only difference is that 
$|\psi_{mk}\rangle$ and $|\psi_{m'k}\rangle$ are no longer statistically independent. 
We thus obtain the following
\begin{theorem}
Select $K$ i.i.d.\ random unitaries $U_{k}$ ($k=1,\dots, K$) from the phase ensemble
and define the codewords $|\psi_{mk}\rangle = U_k |m\rangle$ ($m=1,\dots, d$).
Then, for any $\delta >0$ and
\begin{equation}
K > \frac{4}{\delta^2} \left( \ln{d} + \frac{2}{\delta} \ln{ \frac{5}{\delta} }\right) \, ,
\end{equation}
Eve's accessible information satisfies
\begin{equation}
I_{\text{acc}} = O\left( \delta \log{d} \right)
\end{equation}
up to a probability that vanishes exponentially in $d$.
\end{theorem}
The proof of this theorem can be obtained by a straightforward modification of that of Theorem~\ref{main-r}
and is not reported here.

It is worth noticing that these unitaries form an abelian group. 
It is hence somehow surprising that they yield strong QDL properties.


\section{Sketch of the proof of Theorem~\ref{main-r}}\label{sec:sketch}

To compute Eve's accessible information about the message, we associate to the 
sender Alice a dummy $M$-dimensional quantum system
carrying the classical variable $m$ as a set of basis vectors $\{ |m\rangle \}_{m=1,\dots,M}$. 
We then suppose that Eve intercepts the quantum system that has the encrypted message. 
Since Eve does not know the secret key, the correlations between Alice and Eve 
are described by the following classical-quantum state:
\begin{equation}
\rho_{AE} = \frac{1}{M} \sum_{m=1}^M |m\rangle\langle m|_A \otimes \frac{1}{K} \sum_{k=1}^K |\psi_{mk}\rangle\langle\psi_{mk}|_E \, ,
\end{equation}
where the codewords $|\psi_{mk}\rangle$ are as in Eq.~(\ref{codews}).

The accessible information is by definition the maximum classical mutual
information that can be achieved by local measurements on Alice's and Eve's subsystems:
\begin{eqnarray}
I_{\text{acc}} & := & I_{\text{acc}}(A;E)_{\rho} \nonumber \\
& = & \max_{\mathcal{M}_A,\mathcal{M}_E} H(X) + H(Y) - H(X,Y) \, ,
\end{eqnarray}
where $\mathcal{M}_A \, : A \to X$, $\mathcal{M}_E \, : E \to Y$
are local measurements with output variables $X$ and $Y$ respectively, and
$H(\, \cdot \,)$ denotes the Shannon entropy of the measurement results.

The optimal measurement on Alice's subsystem is obviously a projective measurement on the
basis $\{ |m\rangle \}_{m=1,\dots,M}$. Concerning Eve's measurement, it is sufficient
to consider only rank-one POVM, which are described by measurement operators of the form 
$\{ \mu_j |\phi_j\rangle\langle\phi_j|\}_{j=1\dots,J}$, with $\mu_j \geq 0$, and 
satisfying the normalization condition $\sum_j \mu_j |\phi_j\rangle\langle\phi_j| = \mathbb{I}$.
A straightforward calculation then yields
\begin{multline}
 I_{\text{acc}} = \log{M} 
 -  \\
 \min_{\{\mu_j|\phi_j\rangle\langle\phi_j|\}} \sum_j \frac{\mu_j}{M} \left\{ H[Q(\phi_j)] - \eta\left[ \sum_{m=1}^M Q_m(\phi_j)\right] \right\} \, ,
\end{multline}
where $\eta(x) = -x \log{x}$,
$Q(\phi)$ is the $M$-dimensional real vector of non-negative components
\begin{equation}
Q_m(\phi) = \frac{1}{K} \sum_{k=1}^K |\langle \phi | \psi_{mk} \rangle |^2 \, ,
\end{equation}
and $H[Q(\phi)] = - \sum_{m=1}^M Q_m(\phi) \log{Q_m(\phi)}$ denotes the Shannon entropy of $Q(\phi)$.

We now apply a standard convexity argument, first used in~\cite{QDL}. 
To do that, notice that assuming $\langle \phi_j |\phi_j\rangle = 1$, 
then $\sum_j \mu_j/d = 1$. 
This implies that the positive quantities $\mu_j/d$ can be interpreted as probability weights. 
An upper bound on the accessible information is then obtained by using the fact that
the average cannot exceed the maximum.
We thus obtain
\begin{equation}\label{Iacc-0}
I_{\text{acc}} \leq \log{M} - \frac{d}{M} \min_{|\phi\rangle} \left\{ H[Q(\phi)] - \eta\left[\sum_{m=1}^M Q_m(\phi)\right] \right\} \, .
\end{equation}

Notice that now the maximization is no longer over a POVM with elements $\{ \mu_j |\phi_j\rangle\langle\phi_j| \}$
but over a single normalized vector $|\phi\rangle$.

Then the proof proceeds along three main conceptual steps: 
\begin{enumerate}

\item (See Appendix~\ref{step-1} for details.)
Theorems~\ref{ThRM} and~\ref{ThRM2} imply that the random variable 
$\sum_{m=1}^M Q_m(\phi)$ is close to $M/d$ with arbitrarily high probability for all vectors $|\phi\rangle$
if $d$ is large enough and $M K \gg d$.
In particular, the inequality
\begin{equation}
\sum_{m=1}^M Q_m(\phi) \leq \frac{M}{d} \left[ \left( 1 + \sqrt{\frac{d}{MK}}\right)^2 + \delta \right]
\end{equation}
applied to Eq.~(\ref{Iacc-0}) yields
\begin{equation}
I_{\text{acc}} \leq \alpha \log{d} + \eta(\alpha) - \frac{d}{M} \min_{|\phi\rangle} H[Q(\phi)] \, ,
\end{equation}
where
\begin{equation}
\alpha = \left( 1 + \sqrt{\frac{d}{MK}} \right)^2 + \delta \, .
\end{equation}

\item (See Appendix~\ref{step-2} for details.)
We show that for any given $|\phi\rangle$ and $\delta >0$, 
\begin{equation}
\eta\left[Q_m(\phi)\right] \geq - \frac{1-\delta}{d}\log{\left(\frac{1-\delta}{d}\right)}
\end{equation}
for at least $(1-\delta)M$ values of $m$ (up to a probability exponentially small in $M$).
To do that, we show that there is a negligible probability that $Q_m(\phi) <(1-\delta)/d$ 
for more than $\ell=\delta M$ values of $m$.
This result implies
\begin{equation}\label{Iacc-2-p}
H[Q(\phi)] \geq \frac{M}{d} \left( 1 - 2\delta \right) \log{d} \, .
\end{equation}

\item (See Appendix~\ref{step-3} for details.)
Finally, to account for the minimum over all unit vectors $|\phi\rangle$, we 
introduce a discrete set $\mathcal{N}_\delta$ of vectors 
with the property that for any $|\phi\rangle$ there exists $|\phi_i\rangle \in \mathcal{N}_\delta$
such that 
\begin{equation}
\| |\phi\rangle\langle\phi| - |\phi_i\rangle\langle\phi_i| \|_1 \leq \delta \, .
\end{equation}
A set with this property is called an $\delta$--net. The $\delta$--net is used to
approximate the value of $H[Q(\phi)]$ up to an error of the order of $\delta \log{d}$.
We show that the inequality in~(\ref{Iacc-2-p}) holds true with high
probability for all unit vectors in the $\delta$--net.

\end{enumerate}

In conclusion we obtain that for any $\delta >0$ Eve's accessible information satisfies
\begin{equation}\label{Iacc-3-p}
I_{\text{acc}} = O\left( \delta \log{d}\right) \, , 
\end{equation}
up to a probability which is exponentially small in $d$ provided
\begin{equation}
K  > \frac{4}{\delta^2} \left( \ln{M} + \frac{2d}{\delta M} \ln{\frac{5}{\delta}} \right) \, .
\end{equation}


\section{Robustness against information leakage}\label{sec:robust}

We consider our protocols based on the phase ensemble of random
vectors to assess the robustness of QDL against the leakage 
to Eve of part of the key or the message.

What happens if part of the key is known by Eve?
For example, suppose that Eve knows the first $\gamma \log{K}$ bits 
of the secret key.
Since the remaining $(1-\gamma)\log{K}$ are still secret and
random, it follows that we can still apply Theorem~\ref{main-r} and Eve's accessible 
information satisfies 
$I_\mathrm{acc} = O\left( \delta \log{d} \right)$ up to a failure probability
\begin{equation}\label{pfail1}
p'_\mathrm{fail} \leq \exp{\left[ - M \left( \frac{\delta^3 K^{1-\gamma}}{4} - \delta \ln{M} - \frac{2d}{M}\ln{\frac{5}{\delta}} \right) \right]} \, .
\end{equation}

To assess the security of the QDL protocol under the leakage of any fraction of the key,
we have to take into account all the possible subsets of $\gamma \log{K}$ bits.
Each of these subsets, determines a corresponding subset of $K^{1-\gamma}$ values of the
key that remain secret to Eve. The number of ways in which these values can be chosen is
${ K \choose K^{1-\gamma}}$. 
Applying the union bound we can hence estimate from~(\ref{pfail1}) the failure probability 
if any fraction of $\gamma \log{K}$ bits leaks to Eve:
\begin{equation}
p''_\mathrm{fail} \leq { K \choose K^{1-\gamma}} p'_\mathrm{fail}
\leq \exp{\left( K^{1-\gamma}\ln{K} \right)} \, p'_\mathrm{fail} \, .
\end{equation}
Putting $M/d = \delta$, this probability is exponentially small in $M$ under conditions
\begin{equation}\label{K-cond-1}
K^{1-\gamma} > \frac{4}{\delta^2} \left( \ln{M} + \frac{2}{\delta^2} \ln{\frac{5}{\delta}} \right)
\end{equation}
and
\begin{equation}
M > K^{1-\gamma} \ln{K} \left( \frac{\delta^3 K^{1-\gamma}}{4} - \delta \ln{M} - \frac{2}{\delta} \ln{\frac{5}{\delta}} \right)^{-1} \, .
\end{equation}

In particular, for $\gamma \ll 1$, the condition~(\ref{K-cond-1}) can be
replaced by
\begin{equation}\label{K-cond-2}
K \gtrsim \left[ \frac{4}{\delta^2} \left( \ln{M} + \frac{2}{\delta^2} \ln{\frac{5}{\delta}} \right) \right]^{1+\gamma} \, .
\end{equation}
Compared to~(\ref{K-condition}), this condition implies that to make QDL robust
against the leakage of a fraction of $\gamma\log{K}$ bits of the key, one simply
needs to use a longer key of about $(1+\gamma)\log{K}$ bits.
This result shows the existence of QDL schemes that are robust to some loss of key.


What happens if a small fraction of $n$ bits of the message leaks to Eve?
Suppose that Eve knows the first $n$ bits of the message, then
one has to require that the key is sufficiently long to lock the remaining $\log{M}-n$
bits of the message.
Since the codewords corresponding to the remaining part of the message are still
random, we have
\begin{equation}\label{I-Eve-1}
I_{\text{acc}} = O \left( \delta \left( \log{M} - n \right) \right) \, ,
\end{equation}
up to a probability
\begin{align}
p'_\mathrm{fail} \leq & \exp{\left\{ - M 2^{-n} \left[ \frac{\delta^3 K}{4} - \delta (\ln{M}-n) \right. \right.} \nonumber \\
& \hspace{3cm} \left. \left. - 2^{n+1} \frac{d}{M} \ln{\frac{5}{\delta}} \right] \right\} \, .
\end{align}
We apply again the union bound to estimate the failure probability 
if any subset of $n$ bits of the message leaks to Eve:
\begin{align}
p''_\mathrm{fail} \leq & { M \choose M 2^{-n}} \, p'_\mathrm{fail} \leq \exp{\left(M 2^{-n} \ln{M}\right)} \, p'_\mathrm{fail} \\
\leq & 
\exp{\left\{ - M 2^{-n} \left[ \frac{\delta^3 K}{4} - \ln{M} - \delta (\ln{M}-n) \right. \right. } \nonumber \\
& \hspace{3.5cm} \left. \left. - 2^{n+1} \frac{d}{M} \ln{\frac{5}{\delta}} \right] \right\} \, .
\end{align}
For any given $n$, the latter is exponentially small in $M$ provided
\begin{equation}
K > \frac{4}{\delta^3} \left[ \ln{M} + \delta (\ln{M}-n) + 2^{n+1} \frac{d}{M} \ln{\frac{5}{\delta}} \right] \, .
\end{equation}

Compared to~(\ref{K-condition}), the last condition implies that to protect QDL against the leakage of
$n$ bits of message, the key should be enlarged by $\Delta(\log{K}) \simeq n$ bits.
This result shows the existence of QDL schemes that are robust to plain-text attack.
A similar result can be obtained starting from other QDL protocols, e.g., using
the results of~\cite{Omar,Dupuis} based on the min-entropy of the message.


\section{Applications}\label{sec:appl}

{\it Towards quantum optical realizations.---}
The phase ensemble finds natural applications in the context of linear optics.
For instance, codewords belonging to the phase ensemble can be realized by coherently splitting a
photon over $d$ modes (e.g., spatial, temporal, orbital angular momentum modes, etc.), then by 
applying independent random phase shifts to each mode. 
If information is encoded in the arrival time, the codewords can be prepared by first
applying a linear dispersion transformation (see, e.g.,~\cite{Dirk}) and then random 
phase shifts at different times.

Concerning Bob decoding, in the case of QDL by the phase ensemble of random unitaries
(see Sec.~\ref{ssec:unitary}) it is sufficient to apply the inverse transformation of the one applied
by Alice for encoding, then measure by photo-detection.
Notice that both encoding and decoding operations can be realized by linear passive optical
elements and photodetection (for decoding) as discussed in~\cite{QEM}.
For the QDL based on random codewords from the phase ensemble of vectors, Bob should in principle 
decode by applying the pretty good measurement associated to the set of QDL codewords, 
yet we don't know at the moment an explicit construction.

A crucial requirement to realize our strong QDL protocols in quantum optics
is the ability to prepare and manipulate quantum states of light in high dimension.
The latter is the goal of cutting edge research and several 
important milestone have been achieved so far. 
See, for example, the recent report of entanglement production 
between two photons in a $100 \times 100$-dimensional Hilbert space~\cite{Z}.


{\it Quantum data locking of noisy channels.---} 
Although the phenomenon of QDL has been known for
about ten years now~\cite{QDL}, only recently has it been reconsidered
in the context of noisy quantum channels~\cite{QEM,PRX}.
In particular, the locking capacity of a (noisy) quantum channel has been 
defined in~\cite{PRX} as the maximum rate at which classical information 
can be locked through $N$ instances of the channel
with the assistance of a secret key shared by Alice and Bob which grows
sub-linearly in $N$.
One can indeed define a weak and a strong notion of locking capacity~\cite{PRX}. 
In the weak case one assumes that Eve has access to the 
channel environment (the output of the complementary channel), while in the
strong case one gives her access to the quantum system being input
to the noisy channel.
Remarkably, there are examples of quantum channels whose locking capacity is much
larger than the private capacity~\cite{AW,seplock}.

Our result on Eve's accessible information applies directly
to the strong notion of locking capacity and can be extended
(by a simple application of a data processing inequality) to the weak case.
It hence remains to show how and at which rate Bob can decode reliably.
One way to do that is by concatenating the QDL
protocol with an error correcting code~\cite{Fawzi,Omar,PRX}. 
Consider $N \gg 1$ uses of a quantum channel.
If the quantum capacity of the channel is $Q$, then one can lock information 
by choosing codewords in an error correcting subspace of dimension $d \simeq 2^{N Q}$.
Another approach may consist in designing a code which allows for both 
QDL and error correction at the same time.
Our results indicate that random codes exist that can be applied both for protecting
against decoherence and for locking classical information. (An explicit example has been
recently presented in~\cite{seplock}.)


{\it Locking a quantum memory.---}
The QDL properties of the phase ensemble can be used to lock information into 
a quantum memory by applying a local random gauge field. 
Consider an ideal (noiseless) semiclassical model for a
quantum memory consisting of $d$ charged particles on a ring of length $L$. 
For recording locked information in the quantum memory, Alice applies a 
random i.i.d.~magnetic field to each particle and encodes a classical message 
into one of the momentum eigenstates
$|p\rangle = d^{-1/2} \sum_{k=1}^d e^{i 2\pi p x/L + i \int_0^x A(x')dx'} |x\rangle$,
where $A(x')$ is the vector potential in natural units. 
Notice that the application of the random local fields corresponds to a 
random phase in the momentum eigenstates.
Then, a legitimate receiver who knows the pattern of the magnetic field
applied by Alice, can decode the message by simply measuring the momentum. 
On the other hand, Eve's accessible information can be made negligibly small 
by the QDL effect.


\section{Conclusion}\label{sec:concl}

It is well known that the security provided by QDL can be
severely hampered if even a small fraction of the key or the message leaks to Eve.
Here we show that, although this is true in general, there exist
QDL protocols that are instead robust against the leakage of a small
part of the message or the key.


Until now, the codewords used in QDL have been restricted to 
either Haar-distributed random bases~\cite{CMP,Dupuis,Fawzi} or approximate 
mutually unbiased bases~\cite{Fawzi} (the role of mutually unbiased bases being not
yet completely understood~\cite{QDL,Wehner}).
This paper showed that codewords modulated by random phase shifts in a given basis
suffice to guarantee strong and robust QDL properties.

The fact that random phases suffice to ensure strong QDL properties
yields a major simplification for the experimental realization of a  
QDL protocol. To produce states from the phase ensemble, one only
requires to generate $d$ binary phase shifts, instead of $d^2$ random variables
sampling Haar-distribution of unitary transformations.
The phase ensemble is well-adapted for use in quantum optical channels,
where a single photon may be coherently split across different modes (e.g., path or time-bin modes), 
to which i.i.d.~random phase-shifts are applied. Alternatively, one can employ
random unitaries from a set of dispersive transformations~\cite{Dirk}.
To decode the message, the legitimate receiver can first apply the inverse transformation
of the encoding one (both are linear passive transformations), then measure by photo-detection~\cite{QEM}.

Our results suggest that random codes of the type defined here can be used to perform
direct QDL over noisy quantum channels~\cite{QEM,PRX}, which requires
that the codewords for QDL (encoding for security) also be 
appropriate codewords for combating noise on the channel (encoding
for error correction).
A straightforward way for doing that is to concatenate the QDL
protocol with a quantum error correcting code~\cite{Omar}, allowing for a rate of locked
communication at least equal to the quantum capacity of the channel. 
A fundamental question is whether one can lock information at a rate higher than the
quantum (and private) capacity. Such a question has indeed a positive answer, as shown by
the results recently presented in~\cite{AW,seplock}, providing examples of quantum channels with low
or zero privacy that instead allow for high rate QDL. In particular, the phase ensemble was
originally proposed in~\cite{Lloyd} as a code for attaining the coherent information rate for reliable
quantum communication over a quantum channel (see also~\cite{PSW,Horo,Klesse}). The results proved here
suggest that there exist random codes defined from the phase ensemble that allow for robust
QDL over a noisy, lossy bosonic channel at a rate equal to and possibly larger than the
coherent information.


{\it Acknowledgment.---}
This work was supported by the DARPA Quiness Program through US
Army Research Office award W31P4Q-12-1-0019. 
CL thanks Michele Allegra and Xiaoting Wang for several 
valuable and enjoyable scientific discussions.
The authors are very grateful to Hari Krovi and Saikat Guha 
for their comments and suggestions.


\appendix

\section{Proof of Theorem~\ref{main-r}}\label{app:proof}

We now proceed by describing in details the three main steps of the proof of 
Theorem~\ref{main-r} sketched in Sec.~\ref{sec:sketch}.

\subsection{Random matrix theory}\label{step-1}

We apply Theorems~\ref{ThRM} and~\ref{ThRM2} with $n = MK$ and
\begin{equation}
W = \sum_{m=1}^M \sum_{k=1}^K | \psi_{mk} \rangle \langle e_{mk} | 
\end{equation}
to obtain that
\begin{equation}\label{sandw-Q}
\sum_{m=1}^M Q_m(\phi) \leq \left[ \left( 1 + \sqrt{\frac{d}{MK}} \right)^2 + \delta \right] \frac{M}{d}
\end{equation}
for all unit vectors $|\phi\rangle$, up to a probability exponentially
small in $d$ and $\delta$.


Since $\eta(x) = - x \log{x}$ is a monotonically increasing function near zero and $M/d \ll 1$,  Eq.~(\ref{sandw-Q}) yields
\begin{eqnarray}
\frac{d}{M} \, \eta\left[\sum_{m=1}^M Q_m(\phi) \right] & \leq & \eta \left[ \left( 1 + \sqrt{\frac{d}{MK}} \right)^2 + \delta \right] \\
& - & \left[ \left( 1 + \sqrt{\frac{d}{MK}} \right)^2 + \delta \right] \log{\frac{M}{d}} \, , \nonumber 
\end{eqnarray}
which, applied to~(\ref{Iacc-0}), implies
\begin{equation}\label{Iacc1-d}
I_{\text{acc}} \leq \alpha \log{d} + \eta(\alpha) - \frac{d}{M} \min_{|\phi\rangle} H[Q(\phi)] \, ,
\end{equation}
where
\begin{equation}
\alpha = \left( 1 + \sqrt{\frac{d}{MK}} \right)^2 + \delta \, .
\end{equation}


\subsection{Concentration of the $Q_m$}\label{step-2}

For any given $|\phi\rangle$ and $\delta > 0$, we bound the probability
that there exist $\ell = \delta M \ll M$ values of $m$, say $m_1, m_2, \dots, m_\ell$, 
such that 
\begin{equation}
\eta\left[Q_{m_i}(\phi)\right] < - \frac{1-\delta}{d}\log{\left( \frac{1-\delta}{d} \right)}
\end{equation}
for all $i=1,2,\dots,\ell$.
This accounts for bounding the probability that $Q_m(\phi)$ takes small values
(smaller that $(1-\delta)/d$) as well as the probability of large values 
(larger than $1 - \eta{[(1-\delta)/d]} + O(\eta{[(1-\delta)/d]})$).


{\it Bounding the probability of large $Q_m(\phi)$.} 
We first bound the probability that $Q_m(\phi)$ takes large values.
We apply Theorems~\ref{ThRM} and~\ref{ThRM2} with $n = K$ and $y = d/K$. 
This yields that the inequality 
\begin{equation}
Q_m(\phi) \leq \frac{\left( 1 + \sqrt{d/K} \right)^2 + \delta }{d}
\end{equation}
is satisfied for all unit vectors $|\phi\rangle$, up to a probability
exponentially small in $d$ and $\delta$.
In particular, for $d \gg K$ and $\delta \ll 1$ we obtain
\begin{equation}
Q_m(\phi) \lesssim \frac{1}{K} \, .
\end{equation}


{\it Bounding the probability of small $Q_m(\phi)$.}
To bound the probability that $Q_m(\phi)$ takes small values
we apply the following theorem:

\begin{theorem}\label{ThMaurer}~\cite{AM}
Given $K$ i.i.d.~positive random variables $X_k \sim X$ such that $\mathbb{E}[X]$ and
$\mathbb{E}[X^2]$ are finite, then 
\begin{equation}
\Pr\left\{ \mathbb{E}[X] - \frac{1}{K} \sum_{k=1}^K X_k > \tau \right\} \leq \exp{\left( - \frac{K \tau^2}{2\mathbb{E}[X^2]}  \right)}  \, .
\end{equation}
\end{theorem}

Let us put 
\begin{equation}
q_{mk}(\phi) = |\langle \phi | \psi_{mk} \rangle |^2 = \frac{1}{d} \left| \sum_{\omega=1}^d \phi_\omega e^{-i\theta_{mk}(\omega)} \right|^2 \, .
\end{equation}
We have
\begin{equation}
\mathbb{E}[q_{mk}(\phi)] 
= \frac{1}{d}\sum_{\omega} |\phi_\omega|^2 = \frac{1}{d} \, ,
\end{equation}
and
\begin{eqnarray}
\mathbb{E}[q_{mk}^2(\phi)] 
& = & \frac{1}{d^2} \left[ 2 \sum_{\omega,\omega'} |\phi_\omega|^2 |\phi_{\omega'}|^2 - \sum_\omega |\phi_\omega|^4 \right] \\
& = & \frac{1}{d^2} \left[ 2 - \sum_\omega |\phi_\omega|^4 \right] \, .
\end{eqnarray}
By noticing that $\sum_\omega |\phi_\omega|^4 \geq 1/d$, we obtain
\begin{equation}
\mathbb{E}[q_{mk}^2(\phi)] \leq
\frac{1}{d^2} \frac{2d-1}{d}
\leq \frac{2}{d^2} \, .
\end{equation}

We can thus apply the tail bound in Theorem~\ref{ThMaurer} 
with $X_k = q_{mk}(\phi)$ and $\tau = \delta/d$ 
to obtain
\begin{equation}\label{tailQ2}
\Pr\left\{ Q_m(\phi) < \frac{1-\delta}{d} \right\} 
< \exp{\left( - \frac{K \delta^2}{4}  \right)} \, .
\end{equation}

Having bounded the probability of small values of $Q_m(\phi)$, we now
consider the probability that there exist $\ell$ values of $m$ for which 
$Q_m(\phi) < (1-\delta)/d$.
Taking into account all the ${ M \choose \ell}$ choices of $\ell$ values of
$m = 1, \dots, M$ and applying the union bound we obtain
\begin{align}
& \Pr\left\{ \exists m_1, \dots, m_\ell \, | \, \forall \, j \, , \, Q_ {m_j}(\phi) < \frac{1-\delta}{d} \right\} \nonumber \\
& \hspace{3.2cm} \leq { M \choose \ell} \, \exp{\left( - \frac{\ell K \delta^2}{4}  \right)} \\
& \hspace{3.2cm} \leq \exp{\left( - \frac{\ell K \delta^2}{4}  + \ell \ln{M} \right)} \, . \label{Probsmall}
\end{align}


\subsection{The $\delta$--net}\label{step-3}

It remains to prove that~(\ref{Probsmall}) holds true with high probability
for all unit vectors $|\phi\rangle$.
To show that we follow a standard strategy and introduce an
$\delta$--net of vectors~\cite{CMP}.
An $\delta$--net $\mathcal{N}_\delta$ is a discrete set of $|\mathcal{N}_\delta|$ vectors 
such that for any vector $|\phi\rangle$ there exists $|\phi_i\rangle \in \mathcal{N}_\delta$
such that
\begin{equation}\label{Trd}
\| |\phi\rangle\langle\phi| - |\phi_i\rangle\langle\phi_i| \|_1 \leq \delta \, ,
\end{equation}
where $\| X \|_1 = \mathrm{Tr}|X|$ is the trace norm.


As discussed in~\cite{CMP}, there exist $\delta$--nets in a $d$-dimensional Hilbert 
space with $|\mathcal{N}_\delta| \leq (5/\delta)^{2d}$.
We can then apply the union bound to estimate
\begin{align}
& \Pr\left\{ \exists |\phi_i\rangle \in \mathcal{N}_\delta, m_1, \dots, m_\ell \, | \, \forall \, j \, , \, Q_{m_j}(\phi_i) < \frac{1-\delta}{d} \right\} \nonumber \\
& \hspace{1cm} \leq \left( \frac{5}{\delta} \right)^{2d} \exp{\left[ - M \left( \frac{\delta^3 K}{4} - \delta \ln{M} \right) \right]} \\
& \hspace{1cm} = \exp{\left[ - M \left( \frac{\delta^3 K}{4} - \delta \ln{M} - \frac{2d}{M} \ln{\frac{5}{\delta}} \right) \right]} \, . 
\end{align}
Notice that such a probability is exponentially small in $M$ provided that
\begin{equation}\label{Kcond}
K  > \frac{4}{\delta^2} \left( \ln{M} + \frac{2d}{\delta M} \ln{\frac{5}{\delta}} \right) \, .
\end{equation}

It follows that with probability close to $1$ 
\begin{equation}
\eta\left[Q_m(\phi_i)\right] \geq \frac{1-\delta}{d}\log{d}
\end{equation}
for all $|\phi_i\rangle \in \mathcal{N}_\delta$ and at least $(1-\delta)M$ values of $m$.
This result in turn implies
\begin{align}
H[Q(\phi_i)] 
& \geq \frac{M}{d} \left( 1 - \delta \right)^2 \left[ \log{d} - \log{(1-\delta)} \right] \nonumber \\
& \geq \frac{M}{d} \left( 1 - \delta \right)^2 \log{d} \nonumber \\
& \geq \frac{M}{d} \left( 1 - 2\delta \right) \log{d} \, . \label{H-bound}
\end{align}


It remains to estimate the error introduced by the $\delta$--net.
The Fannes-Audenaert~\cite{FA} inequality yields
\begin{eqnarray}
\left| H(Q(\phi)) -  H(Q(\phi_i)) \right| & \leq & \frac{\| Q(\phi) -  Q(\phi_i) \|_1}{2} \log{M} \nonumber  \\
& + & h_2\left[\frac{\| Q(\phi) -  Q(\phi_i) \|_1}{2} \right] \, , \nonumber
\end{eqnarray}
where $\| Q(\phi) -  Q(\phi_i) \|_1 = \sum_m | Q_m(\phi) -  Q_m(\phi_i) |$
denotes the vector trace norm
and $h_2(x) = - x \log{x} - (1-x)\log{(1-x)}$ is the binary entropy.
It is shown in Appendix~\ref{app:norms} that Eq.~(\ref{Trd}) implies
\begin{equation}
\| Q(\phi) - Q(\phi_i) \|_1 \leq 2 \delta \frac{M}{d} \, ,
\end{equation}
up to a probability exponentially small in $d$ and $\delta$.


\subsection{Conclusion}

Putting all this together, we obtain that if Eq.~(\ref{Kcond}) is verified then 
\begin{eqnarray}
I_{\text{acc}} & \leq & \left( \alpha - 1 + 2\delta\right) \log{d} \nonumber \\
& + & \eta(\alpha) + \delta M/d \log{M} + h_2(\delta M/d) \\
& \leq & \left( \alpha - 1 + 3\delta\right) \log{d} + \eta(\alpha) + h_2(\delta) \, , \label{Iacc-3}
\end{eqnarray}
up to a probability exponentially small in $M$.
Equation~(\ref{Kcond}) implies
\begin{equation}
K > \frac{1}{\delta^3} \, \frac{d}{M} \, ,
\end{equation}
which in turn yields
\begin{equation}
\alpha - 1 < \delta + 2 \delta^{3/2} + \delta^3 \, .
\end{equation}

In conclusion we have that for any $\delta$ and $d$ large enough
there exist QDL protocols defined from the phase ensemble such that
\begin{equation}
I_{\text{acc}} = O\left( \delta \log{d} \right) \,,
\end{equation}
with a pre-shared secret key of $\log{K}$ bits and
\begin{equation}
K  > \frac{4}{\delta^2} \left( \ln{M} + \frac{2d}{\delta M} \ln{\frac{5}{\delta}} \right) \, .
\end{equation}


\section{Relation between trace norms}\label{app:norms}

To find an explicit relation between the trace norms 
$\| |\phi\rangle\langle\phi| - |\phi_i\rangle\langle\phi_i| \|_1$
and $\| Q(\phi) - Q(\phi_i) \|_1$, we introduce the operators
\begin{equation}
\Gamma_m = \frac{d}{M} \left[ \left( 1 + \sqrt{\frac{d}{MK}} \right)^2 + \delta \right]^{-1} \frac{1}{K} \sum_k |\psi_{mk}\rangle\langle\psi_{mk}| \, .
\end{equation}

Notice that the operators $\{ \Gamma_m \}_{m=1,\dots,M}$ are positive, and
Theorems~\ref{ThRM} and~\ref{ThRM2} imply that $\sum_m \Gamma_m \leq \mathbb{I}$, 
up to a probability exponentially small in $d$ and $\delta$.
They hence define an incomplete POVM.

For any unit vector $|\phi\rangle$, the output of this incomplete POVM
is the subnormalized probability vector $\tilde Q(\phi)$ with entries
\begin{equation}
\tilde Q_m(\phi) = \langle \phi | \Gamma_m | \phi \rangle = 
\frac{d}{M} \left[ \left( 1 + \sqrt{\frac{d}{MK}} \right)^2 + \delta \right]^{-1} \hspace{-0.3cm} Q_m(\phi) \, .
\end{equation}

Since an incomplete POVM does not increase the trace distance, we have that if
\begin{equation}
\| |\phi\rangle\langle\phi| - |\phi_i\rangle\langle\phi_i| \|_1 \leq \delta
\end{equation}
then
\begin{equation}
\| \tilde Q(\phi) - \tilde Q(\phi_i) \|_1 \leq \delta \, ,
\end{equation}
which in turn implies
\begin{equation}
\| Q(\phi) - Q(\phi_i) \|_1 \leq \delta \frac{M}{d} \left[ \left( 1 + \sqrt{\frac{d}{MK}} \right)^2 + \delta \right] \, .
\end{equation}
For sufficiently small $\delta$ and $d/(MK)$ this implies
\begin{equation}
\| Q(\phi) - Q(\phi_i) \|_1 \leq 2 \delta \frac{M}{d} \, .
\end{equation}


\begin{thebibliography}{99}
\bibitem{NOTA-design}
As noticed in~\cite{Fawzi,Omar}, unitary designs~\cite{design} are not suitable for defining
strong QDL schemes. In particular, they don't allow in general for a secret key
of constant size. We think the same holds true for local random circuits~\cite{a-design}.
\bibitem{NOTA-coding}
Random coding is a standard technique to cope with noisy information transmission.
QDL represents no exception, as well as other protocols related to QDL where coding
techniques have been already applied, e.g., quantum hiding fingerprint~\cite{finger} and
the bounded-quantum-storage model for quantum key distribution~\cite{bound-model}.
In particular, codewords of the form~(\ref{phase-v}) were also considered in~\cite{finger} 
in the context of quantum hiding fingerprints.
\bibitem{NOTA1}
An alternative
criterion for QDL is obtained in terms
of the total variational distance. Suppose that Eve applies a
measurement with outcomes labeled by the index $l$. Then, for any of the messages $m$
one has a conditional probability $p(l|m)$ for obtaining measurement
outcome $l$ when message $m$ is transmitted. The total variational distance criterion for 
QDL requires
\begin{equation}\label{eq:TV-criterion}
D = \sum_{l,m} \left| p(l|m)p(m) - p(l) p(m) \right|
\end{equation}
to be small for all measurements of Eve,
where $p(m)$ is the probability of transmitting message $m$ and $p(l) = \sum_m p(l|m) p(m)$. 
The interpretation here is that if~(\ref{eq:TV-criterion}) is small,
then Eve cannot do much better than random guessing to determine the message. 
As discussed in~\cite{Fawzi}, a straightforward application of the Alicki-Fannes inequality~\cite{Fannes} 
implies that if $D \leq \delta/2$, then $I_{\text{acc}} \leq 4 \delta \log{d} + 2h_2(\delta)$,
where $h_2(\delta) = -\delta\log{\delta}-(1-\delta)\log{(1-\delta)}$ is the binary entropy.
On the other hand, the Pinsker inequality implies that (see e.g.,~\cite{Pisker} and references therein) 
\begin{equation}
D \leq  \sqrt{(2 \ln{2})I_{\text{acc}}} \, .
\end{equation}
That is, if $I_{\text{acc}} \leq \delta \log{d}$ then $D \leq \sqrt{2 \delta \ln{d}}$.
\end{thebibliography}
\end{document}